# Actively controlling the topological transition of dispersion based on electrically controllable metamaterials

Zhiwei Guo, Haitao Jiang*, Yong Sun, Yunhui Li, and Hong Chen

*Key Laboratory of Advanced Micro-structure Materials, MOE, School of Physics Science and Engineering, Tongji*

*University, Shanghai 200092, China*

**Abstract**

Topological transition of the iso-frequency contour (IFC) from a closed ellipsoid to an open hyperboloid, will provide unique capabilities for controlling the propagation of light. However, the ability to actively tune these effects remains elusive and the related experimental observations are highly desirable. Here, tunable electric IFC in periodic structure which is composed of graphene/dielectric multilayers is investigated by tuning the chemical potential of graphene layer. Specially, we present the actively controlled transportation in two kinds of anisotropic zero-index media containing PEC/PMC impurities. At last, by adding variable capacitance diodes into two-dimensional transmission-line system, we present the experimental demonstration of the actively controlled magnetic topological transition of dispersion based on electrically controllable metamaterials. With the increase of voltage, we measure the different emission patterns from a point source inside the structure and observe the phase-transition process of IFCs. The realization of actively tuned topological transition will opens up a new avenue in the dynamical control of metamaterials.

* Corresponding author: Email: jiang-haitao@tongji.edu

# I. INTRODUCTION

Recently, hyperbolic metamaterials (HMMs) have greatly attracted people's attention for its open iso-frequency contour (IFC) in the wave vector space. Once the topological transition have been realized from a closed ellipsoid to an open hyperboloid, interaction between light and matter will be significant changed, such as enhanced spontaneous emission [1-4], all-angle negative refraction [5-9], super-resolution imaging [10-14], long-range interaction [15-17], Cherenkov emission with low energy electrons [18], etc. Up to now, by varying the frequency, the sign of the real part of permittivity ($\varepsilon$) or permeability ($\mu$) in dispersive metamaterials is changed and thereby the topological transition of dispersion is realized [1, 8, 19, 20]. In addition, by maintaining the real parts and tuning the imaginary part of $\varepsilon$ or $\mu$ at a fixed frequency, topological transition of IFC have been theoretically analyzed [21] and experimentally demonstrated in metamaterials with added losses [22, 23]. In contrast to the passive metamaterials, the study of actively tuned metamaterials and meta-devices is also an outstanding research topic [24-27]. Actively controlled metamaterial systems have been predicted to be able to yield new applications ranging from electrically switchable devices [26] to the tunable coupling devices [25, 27]. The realization of the actively tunable topological transition of IFC is also very useful in the design of new active optical devices. To realize the externally tunable behaviors, two-dimensional (2D) semiconductors such as graphene are usually utilized. Graphene, a 2D honeycomb lattice of carbon atoms, is electrically characterized by its surface conductivity $\sigma(\omega, \mu_c)$, where $\mu_c$ is the chemical potential that can be tuned by the gate voltage [24], and $\omega$ denotes the angular frequency. Tunable absorption [28, 29], giant Purcell effect [30] and hyperbolic plasmonics [31, 32] have been theoretically proposed in graphene-based structures associated with the hyperbolic property at the suitable condition. In this paper, by tuning the chemical potential of graphene layer, we theoretical analyze and numerically simulate the tunable emission patterns in periodic structure composed of graphene/dielectric multilayers. The emission patterns clearly show the topological transition of IFC from a closed ellipsoid to an open hyperboloid [19, 23]. Remarkably, by actively changing the chemical potential of graphene layer, two special cases corresponding to two kinds of anisotropic $\varepsilon$-near-zero (ENZ) media can be realized in the proposed structure at a fixed frequency. To see the influence of the topological transition on the wave propagations, we study the transmission property of the two kinds of ENZ media embedded with a PEC or PMC defect. Furthermore, we propose a microwave experimental platform to demonstrate the actively controlled topological transition of IFC. By adding electrically controlled diodes into 2D transmission-line (TLs) system, we experimentally observe the actively tuned topological transition from the emission patterns by modulating the external voltages applied in the diodes.

The paper is organized as follows. In Sec. II, by tuning the chemical potential of graphene layer, we show an actively adjustable topological transition of dispersion in the periodic structure composed of graphene and dielectric layers. Moreover, the influences of the topological transition of dispersion on the scattering properties of structures with embedded PEC or PMC defect have been fully studied. Then, in Sec. III, based on a 2D TL system, we experimentally demonstrate the actively controlled topological transition of dispersion based on electrically controllable metamaterials. Finally, we conclude in Sec. V.

## II. ACTIVELY TUNABLE TOPOLOGICAL TRANSITION IN GRAPHENE/DIELECTRIC MULTILAYERS

We consider the structure, shown in Fig. 1, which is composed of graphene layers separated by dielectric slabs. For the graphene sheet, electromagnetic properties are characterized by its surface conductivity $\sigma(\omega, \mu_c)$. The $\sigma(\omega, \mu_c)$ can be calculated as the Kubo formula [33, 34]:

$$\sigma(\omega, \mu_c) = \sigma_{intra}(\omega, \mu_c) + \sigma_{inter}(\omega, \mu_c), \quad (1)$$

$\sigma_{intra}(\omega, \mu_c)$ and $\sigma_{inter}(\omega, \mu_c)$ are denote the conductivity due to the intra-band and inter-band contribution respectively.

$$\begin{aligned}\sigma_{intra}(\omega, \mu_c) &= \frac{-ie^2}{\pi\hbar^2(\omega + i\tau^{-1})}[\int_0^\infty p(\frac{\partial f_d(p)}{\partial p} - \frac{\partial f_d(-p)}{\partial p})dp], \\ \sigma_{inter}(\omega, \mu_c) &= \frac{ie^2(\omega + i\Gamma)}{\pi\hbar^2}[\int_0^\infty \frac{f_d(-p) - f_d(p)}{(\omega + i\tau^{-1})^2 - 4(p/\hbar)^2}dp]\end{aligned} \quad (2)$$

where $f_d(p)$ is the Fermi-Dirac distribution function: $f_c(p) = \{1 + \exp[(p - \mu_c)/(k_B T)]\}^{-1}$, $k_B, \omega, \hbar, e, T$ and $\tau^{-1}$ is the Boltzmann's constant, angular frequency, Planck constant, charge of an electron, temperature and collision rate, respectively. $\tau^{-1} = 10^{12}$ is assumed independent on the energy $p$. $\mu_c$ is the chemical potential that can be tuned by the external voltage $v_g$ [24]. At high frequencies, the contribution from inter-band can be neglected. For the Fermi-Dirac statistics, conductivity can be simplified as [33]:

$$\sigma(\omega, \mu_c) = \frac{ie^2 \mu_c}{\pi \hbar^2 (\omega + i\tau^{-1})}. \tag{3}$$

When electronic band structure of a graphene sheet is unaffected by the neighboring, the effective permittivity of graphene can be calculated as follow [24]:

$$\varepsilon_g = 1 + i\sigma(\omega, \mu_c) / \varepsilon_0 \omega t_g, \tag{4}$$

where $t_g = 1nm$ and $\varepsilon_0$ are the thickness of graphene and permittivity of vacuum, respectively. For the varied chemical potential $\mu_c$, the different effective complex permittivity of graphene based on the Eq. (4) are shown in Fig. 2(a). The real and imaginary part of $\varepsilon_g$ are marked by solid and dashed lines, respectively. We can clearly find that the complex value of $\varepsilon_g$ is sensitive to the value of $\mu_c$, and the real part of $\varepsilon_g$ will red shift with the $\mu_c$ increased. Owning to the flexibly regulated $\varepsilon_g$, the active TTD can be easily realized. Considering the graphene/dielectric multilayer structure, the components of the effective dielectric permittivity tensor parallel ($\varepsilon_{//}$) and perpendicular ($\varepsilon_\perp$) to the anisotropic axis are given by the effective medium theory [35]

$$\varepsilon_{//} = \frac{t_g + t_d}{t_g / \varepsilon_g + t_c / \varepsilon_c}, \varepsilon_\perp = \frac{\varepsilon_g \cdot t_g + \varepsilon_d \cdot t_d}{t_g + t_d}, \tag{5}$$

where $t_d = 10nm$ and $\varepsilon_d = 4$ are the thickness and permittivity of dielectric layer for hexagonal boron-nitride (h-BN) [36]. According to Eq. (5), the value and the sign of $\varepsilon_{//}$ can be flexible adjusted by changing the value of $\mu_c$, which is shown in Fig. 2(b). Considering a fixed frequency $f = 385THz$, when the value of $\mu_c$ increase from 0.1e to 0.5e, the sign of $\varepsilon_{//}$ reversal at the critical point 0.439e while the sign of $\varepsilon_\perp$ remain positive, as displayed in Fig. 3(a).

Considering the light with TM polarization propagating in the x-y plane of the 2D uniaxial media, the iso-frequency surface in such a strongly anisotropic metamaterial is given by [12]:

$$\frac{k_x^2 + k_y^2}{\varepsilon_{//}} + \frac{k_z^2}{\varepsilon_\perp} = \mu \cdot k_0^2, \qquad (6)$$

where $k_x$, $k_y$ and $k_z$ are the $x$, $y$ and $z$ components of the wave vector respectively. $k_0 = \omega/c$ is the vector in free space. $c$ is speed of light and $\mu$ is the effective magnetic permeability. The IFC in the wave-vector space ($k_x - k_z$) can be obtained according to Eq. (6). In the low chemical potential case $\mu_c = 0.1e$, at $f = 385THz$, the IFC is a closed ellipsoid, which is shown by red line in Fig. 3(b). The solid (dash) lines give the relationship of Re($k_z$)(Im($k_z$)) with $k_x$. By using COMSOL MULTI-PHYSICS based on the finite-element method, the emission pattern to the ellipsoid IFC is shown in Fig 3(c). In the simulation, a current source is put on the center of the material and the perfect matched layer around the material is used to avoid the boundary reflection. However, when the chemical potential is changed to $\mu_c = 0.5e$, the IFC of the graphene/dielectric multilayers will be a hyperbola, which is marked by the solid blue line in Fig. 3(b). The change of IFC will strongly modify the propagation and emission properties of electromagnetic waves. The emission pattern is an open line because of the density of states is maximum along the hyperbolic asymptote (Fig. 3(d)) [37]. The chemical potential of graphene is modulated by an applied electric field and the topological transition of dispersion is realized with the changed chemical potential [24-32]. This actively controlled topological transition will greatly change the emission pattern of a source in the medium. From the simulated emission patterns, we can find that the emission patterns as shown in Fig. 3(c) and 3(d) coincide with the IFCs based on the effective medium theory (Fig. 3(b)). So, the actively tunable IFCs in graphene/dielectric multilayers are realized by tuning the chemical potential of graphene layer.

We further study the influences of the topological transition of dispersion on the scattering properties of structures if PEC or PMC defect is added. According to the Eq. (5), the sign of $\varepsilon_\perp$ also can be controlled with the varied $\mu_c$, which is shown in Fig. 4(a). The effective parameters of $\varepsilon_\perp$ and $\varepsilon_{//}$ are marked by green dashed line and black line, respectively. We can clearly see that the sign of $\varepsilon_\perp$ will change at a critical point 18e as the value of $\mu_c$ increases. At the critical state $\mu_c = 18e$, the IFC is a very flat elliptic curve along horizontal direction (marked by orange line in Fig. 4(b)), which can be regarded as one kind of anisotropic ENZ medium. As another kind of anisotropic ENZ medium for $\mu_c = 0.439e$ in Fig. 3(a), the IFC is a very flat elliptic curve along vertical direction (marked by purple line in Fig. 4(b)). So far, based on passive system, unusual transportation properties of the light in anisotropic ENZ metamaterials have been demonstrated, including collimation [38], flux manipulation [39] and total transmission (reflection) [40]. Now, with the aid of active control, we study the actively tuned transmissions of two kinds of anisotropic ENZ media based on the graphene/dielectric multilayers. For the type Ⅰ of anisotropic ENZ medium ($\mu_c = 18e$), when a plane wave is incident on this structure embedded with a tiny PEC defect, the wave can perfectly pass through this defect without any influence, as is shown in Fig.4 (c). The transmission behavior will change once the PEC defect is replaced by a PMC defect and the incident wave will be scattered by the defect (Fig. 4(d)). Similar to the type Ⅰ of anisotropic ENZ medium, for the type Ⅱ of anisotropic ENZ medium ($\mu_c = 0.439e$), Fig. 4(e) displays that the incident wave also will not be affected by the PEC defect just as in the case of Fig. 4(c). However, the transmission will dramatically change for the structure containing PMC defect. In this case, the incident wave will be completely blocked by the tiny defect (Fig.4 (f)), which is just as the total reflection effect realized in double near-zero-index material [41]. So, inspired by the demonstration of the topological transition of dispersion, we present the novel transportation behavior that can be controlled in an active manner.

## III. EXPERIMENTAL DEMONSTRATION OF THE ACTIVELY TUNABLE TOPOLOGICAL TRANSITION BASED ON TLS

In visible range, the experimental realization of actively tunable topological transition of dispersion is still a great challenge. In this section, we introduce a microwave platform based on 2D TLs to experimentally demonstrate the actively controlled topological transition. The experimental schematic of the TLs-based metamaterial is shown in Fig. 5(a) and the corresponding circuit model is shown in Fig. 5(b). The insets below Fig. 5(b) show the enlarged lumped variable capacitance diodes and the protected elements, respectively. Figure 5(c) shows the schematic of the experimental platform to realize the actively control topological transition. For better visibility, the number of the unit cell ($8 \times 8$) in Fig. 5(c) is smaller than the experimental sample ($20 \times 20$) in Fig. 5(a). In the experiment, a direct voltage source is connected to the sample from the top, which is marked by the red arrow. The grounding position is indicated by red dots. The signal is input at the center of the sample and the protection component is added to avoid the interaction between direct-current (DC) source and the signal source, as is shown in Fig. 5(c).

Here, our sample is composed of $20 \times 20$ unit cells with lumped elements that are fabricated on a commercial printed circuit board, F4B, with a relative permittivity of $\varepsilon_r = 2.2$ and thickness of $h = 1.6 mm$. The width of the micro-strip lines is $w = 2mm$. The unit lengths of TL in the x and z direction are $d = 12mm$. Structural factor of the TL is defined as $g = Z_0 / \eta_{eff}$, where $Z_0$ and $\eta_{eff}$ are the characteristic impedance and the effective wave impedance of the normal TL, respectively. When $w > h$, the structural factor $g = [1.393 + w/h + 0.667 \ln(w/h + 1.444)]^{-1} \approx 0.303$. Then, we load elements into the normal TL to realize the tunable metamaterials. Considering the transverse-electric (TE) wave ($E_y, H_x, H_z$), According to the quasi-static TE polarized solution and Ampere's law, by loading the variable capacitance diode $C$, the effective permittivity and permeability of 2D TLs in the long-wavelength limit can be written as [22, 42]:

$$\begin{aligned}
\varepsilon &= 2C_0 \cdot g / \varepsilon_0, \mu_x = L_0 / (p \cdot \mu_0), \\
\mu_z &= (\frac{L_0}{g} - \frac{1}{\omega^2 \cdot C \cdot d \cdot g}) / \mu_0,
\end{aligned} \quad (7)$$

where $\mu_0$ is permeability of vacuum. $L_0$ and $C_0$ are the per-unit-length inductance and capacitance of TL segment, respectively. The effective permittivity $\varepsilon \approx 3.57$ is isotropic in our structure. For the varactors, one can change the external voltage to tune the capacitance value of $C$. The reference relationship between bias voltage and the diode capacitance provided by the product manual is indicated by the blue dotted line in Fig. 6. From Eq. (7), by continuously changing the value of $C$, the sign and value of $\mu_z$ can be actively controlled at a fixed frequency. Through the modulation of capacitance, we can manipulate the sign of $\mu_z$ from negative to positive while the sign of $\varepsilon$ and $\mu_x$ maintain positive, which is shown in Fig. 7.

The 2D TLs can be described by the effective medium theory and the dispersion relation is derived as [17]:

$$\frac{k_x^2}{\mu_z} + \frac{k_z^2}{\mu_x} = \varepsilon \cdot k_0^2. \tag{8}$$

Based on Eq. (8), we calculate different IFCs by varying the value of $C$ at 0.8 GHz, as is shown in Fig. 8(a). One can see that the topological transition of IFC will happen once the value of $C$ changes from 10 pF to 4 pF. Moreover, the IFC will become more and more flat with the decrease of $C$.

We further use the CST (computer simulation technology) microwave studio software to perform the simulation. A linearly polarized source is loaded near the center of the sample marked by the blue point. In Fig. 8(b), we simulated $E_y$ patterns of sample at 0.8 GHz by varying the value of $C$. The emission patterns for different values of $C$ coincide well with the IFCs in Fig. 8(a). For example, for $C=10$ pF, the IFC is a closed ellipsoid and the source can propagate in all in-plane directions. However, for $C=4.0$ pF, the IFC is an open hyperbolic curve and the source can only propagate in some directions. In particular, for $C=1.0$ pF, the IFC is very flat. The source is collimated and can only propagate in the vertical direction.

Now we perform the experiment to observe the emission patterns. In the experiment, a DC voltage source is connected to the sample form the top, which is marked by the red arrow in Fig. 5. Philips BB181 is used as the variable capacitance diode (marked by green color for see). In order to compare with the simulations, we take a series of voltage values, 1 V, 5 V, 10 V, 15 V, 20 V, and 25V, respectively. The experimental date about the relationship between the used voltage and the diode capacitance is displayed by the red stars in Fig. 6. It is seen that the experimental data coincide well with the reference data. Under the proper design of the shunt connection, the voltage of all the variable capacitance diodes is nearly the same. The signal is emitted from the port one of vector network analyzer (Agilent PNA Network Analyzer N5222A) and the electric fields $E_y$ are measured by another antenna (i.e., near-filed probe) connecting to the port 2 of the network analyzer. The near-field probe is vertically placed 1 mm above the TLs to measure the signals of electric fields of the TLs. In order to accurately probe the field distributions in the near-field scanning measurement, our experimental sample is placed on an automatic translation device. The spatial step of scanning the near field is set to be 1 mm in the x and y directions, respectively. With the increase of the voltage, we measure the different emission patterns at 0.8 GHz to observe the changing process of IFCs. The measured $E_y$ patterns in Fig. 8(c) agree well with the numerical simulated one in Fig. 8(b). So, based on 2D TLs, we have experimentally demonstrated the actively controlled magnetic topological transition of dispersion by changing the external voltages.

## IV. CONCLUTION

In conclusion, we demonstrate the actively tuned topological transition of dispersion in graphene/dielectric multilayers and the influence on the scattering properties when a PEC or PMC defect is embedded in the structure. Moreover, based on 2D TLs loaded with lumped variable capacitance diodes, we experimentally realize the actively controlled topological transition of dispersion. Our results represent a step towards the active control of wave propagations based on metamaterials.

## ACKNOWLEDGMENT

This work is supported by the National Key R&D Program of China (No. 2016YFA0301101); by the National Nature Science Foundation of China (NSFC) (Grant Nos. 11774261, 11474220, and 61621001); Science Foundation of Shanghai (No. 17ZR1443800).

**Figure Captions**

FIG. 1. (Color online) The scheme of the periodic structure composed of graphene/dielectric multilayers. The chemical potential of graphene layer can be tuned by the external voltage. The dielectric layers (h-BN) are marked by yellow and the graphene layers are shown by the single-layer of carbon atoms.

FIG. 2. (Color online) (a) Different effective complex permittivity of graphene. The real part and imaginary part of $\varepsilon_g$ are marked by solid and dashed lines, respectively. (b) The effective parallel permittivity tensor $\varepsilon_{//}$ can be tuned by the different.

FIG. 3. (Color online) (a) Considering a fixed frequency $f = 385 THz$, the effective parameters $\varepsilon_{//}$ and $\varepsilon_{\perp}$ can be tuned by the different $\mu_c$ based on the multilayers structure. As the value of $\mu_c$ increases from 0.1e to 0.5e, the sign of $\varepsilon_{//}$ reversal and the sign of $\varepsilon_{\perp}$ remains positive. (b) When $\mu_c = 0.1e$, the IFC is a closed ellipsoid. While the IFC will become an open hyperbola when $\mu_c = 0.5e$. The solid (dash) lines give the relationship of $\text{Re}(k_z)(\text{Im}(k_z))$ with $k_x$. (c) The emission patterns to the ellipsoid IFC ($\mu_c = 0.1e$) in the graphene/dielectric multilayers. (d) Same as in (c), but obtained for the hyperbola IFC ($\mu_c = 0.5e$).

FIG. 4. (Color online) (a) When the value of $\mu_c$ increases from 15e to 21e, the sign of $\varepsilon_{\perp}$ is reversed and the sign of $\varepsilon_{//}$ remains positive. (b) When $\mu_c = 18e$, the IFC is a very flat elliptic curve, which can be regarded as a highly anisotropic ENZ medium. Distinct from this case, the IFC corresponding to the case of $\mu_c = 0.439e$ is another flat elliptic along vertical

direction. (c) When a plane wave is incident on the structure ($\mu_c = 18e$) with a tiny PEC defect, the wave can perfectly pass through the structure almost without affected by the defect. (d) The transmission behavior will change once the PEC defect is replaced by a PMC defect and the incident wave will be scattered by the defect. (e) For $\mu_c = 0.439e$, the incident wave also will be affected by the PEC defect just as the case of (c). However, the transmission will dramatically change for the PMC defect. (f) For $\mu_c = 0.439e$, the incident wave will be completely blocked by the tiny defect.

FIG. 5. (Color online) (a) The experimental schematic of the TL-based actively controlled metamaterial. The insets show the enlarged lumped variable capacitance diodes and protected elements, respectively. (b) The related anisotropic 2D-circuit model of microwave media. (c) Schematic of part of the structure to realize an actively controlled topological transition. A direct voltage source is connected to the sample from the top, which is marked by the red arrow.

FIG. 6. (Color online) Diode capacitance as a function of bias voltage. Blue dotted line and the red stars are the reference data and experimental data, respectively.

FIG. 7. (Color online) The variation of the effective parameters of TLs with the capacitances of varactors.

FIG. 8. (Color online) The IFCs (a), the corresponding simulated normalized $E_y$ patterns (b) for the different capacitances, and the measured normalized $E_y$ patterns (c) for the different corresponding external voltages.

# Figures

## FIG.1

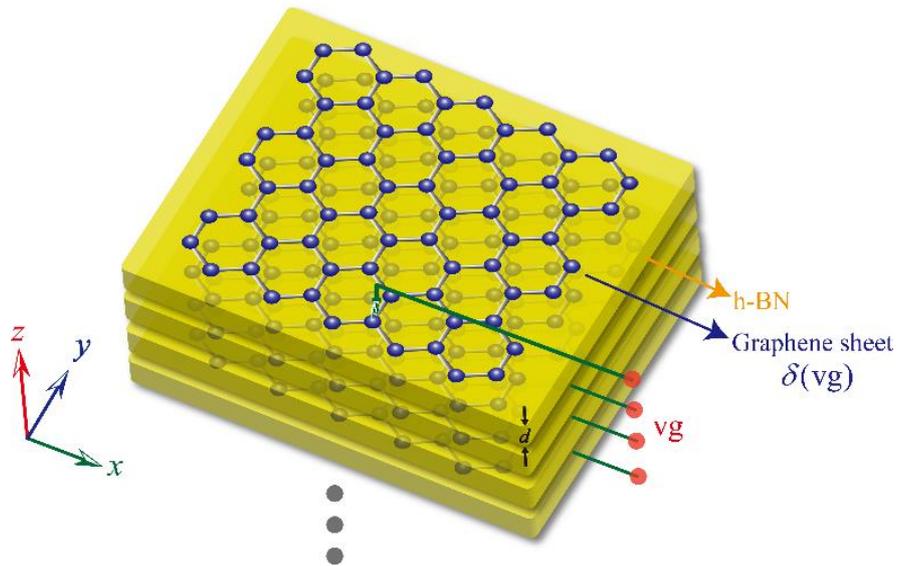

FIG.2

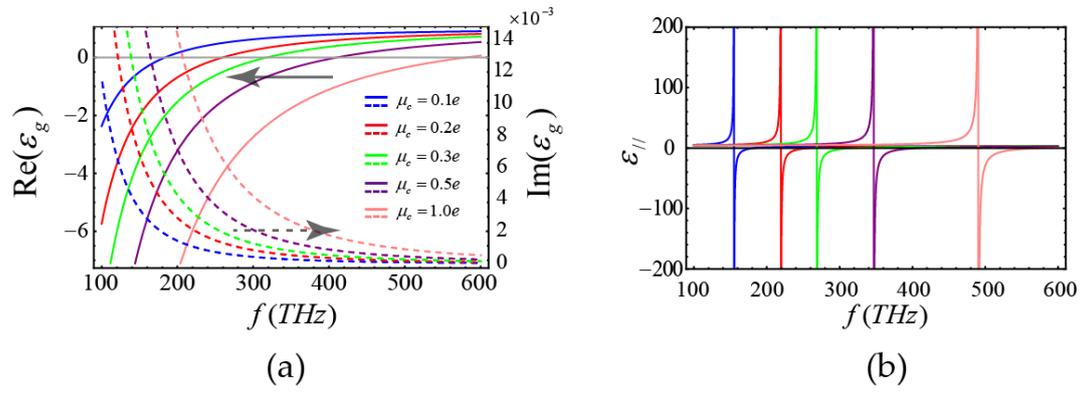

(a)  (b)

FIG.3

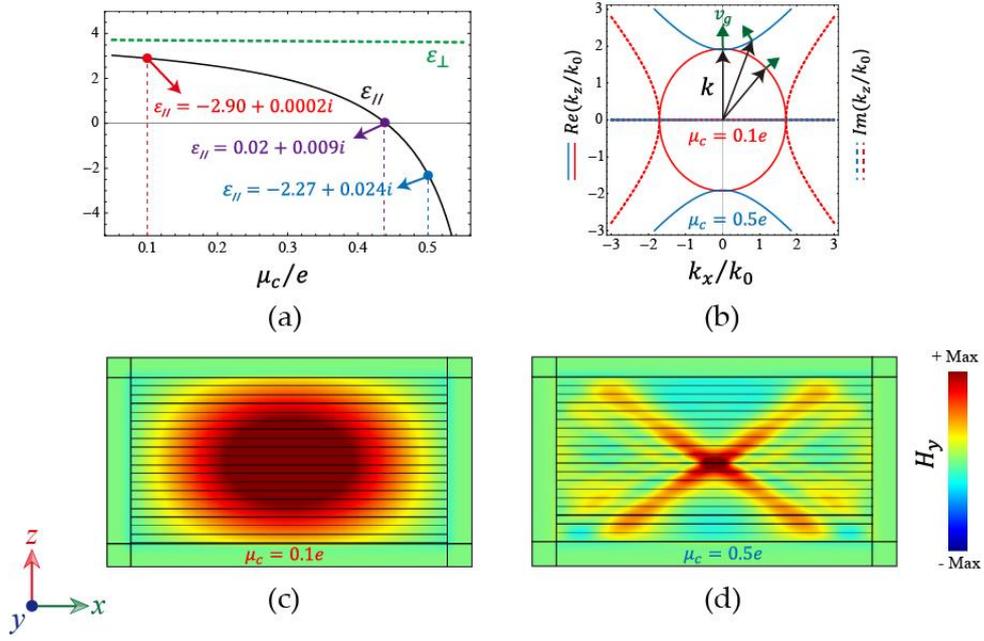

FIG.4

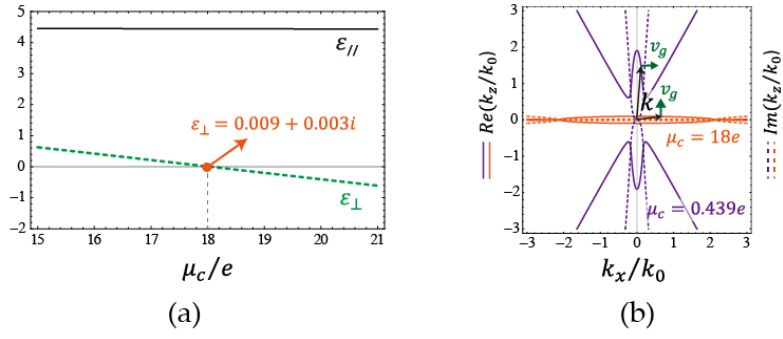

(a)           (b)

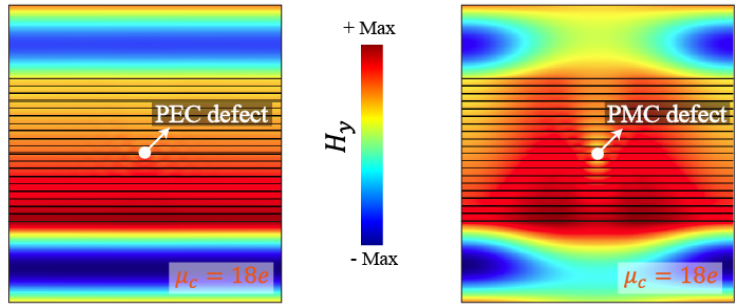

(c)           (d)

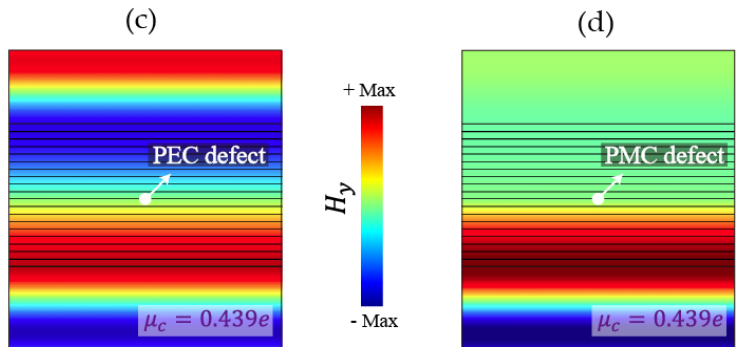

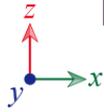

(e)           (f)

FIG.5

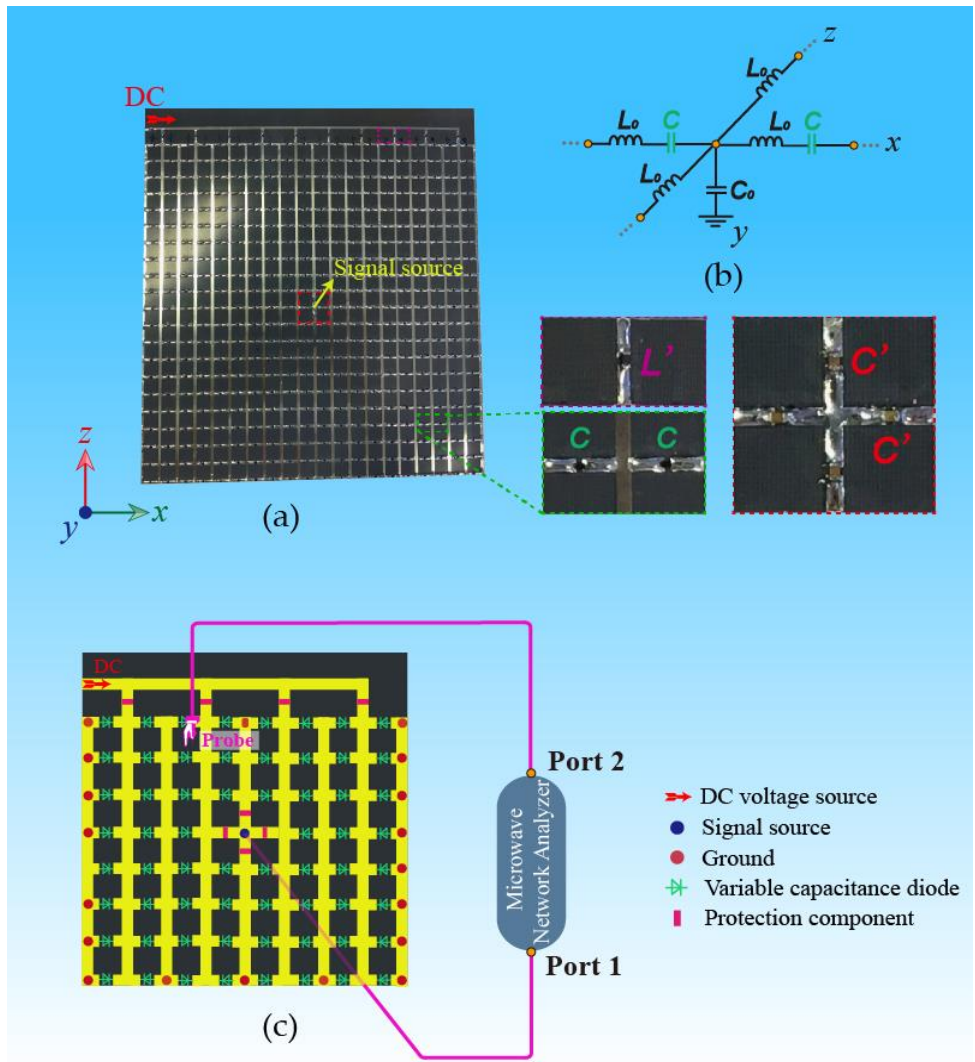

FIG.6

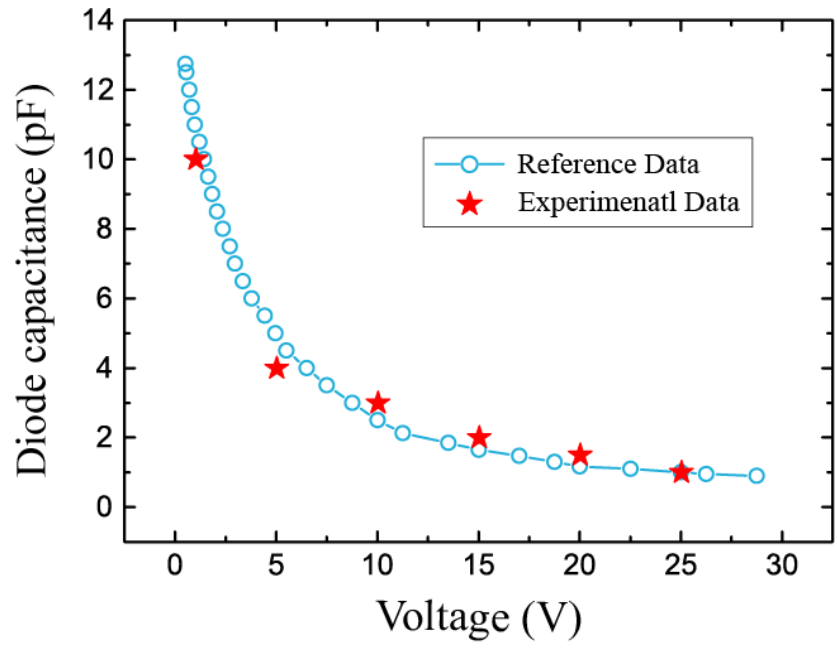

FIG.7

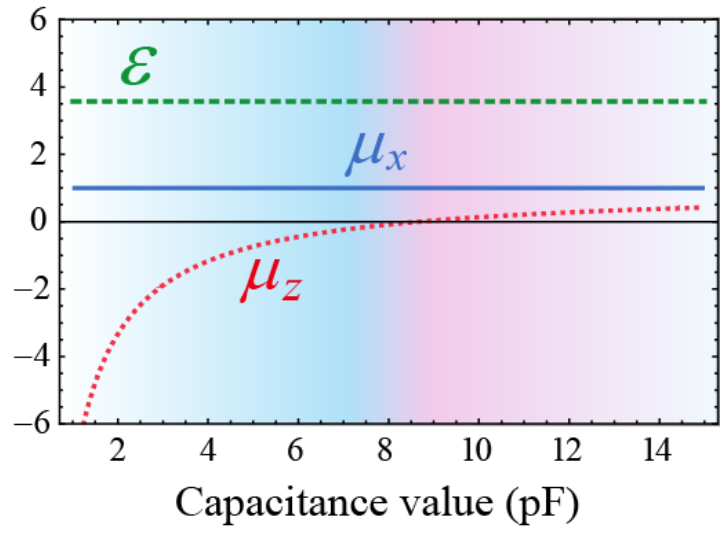

FIG.8

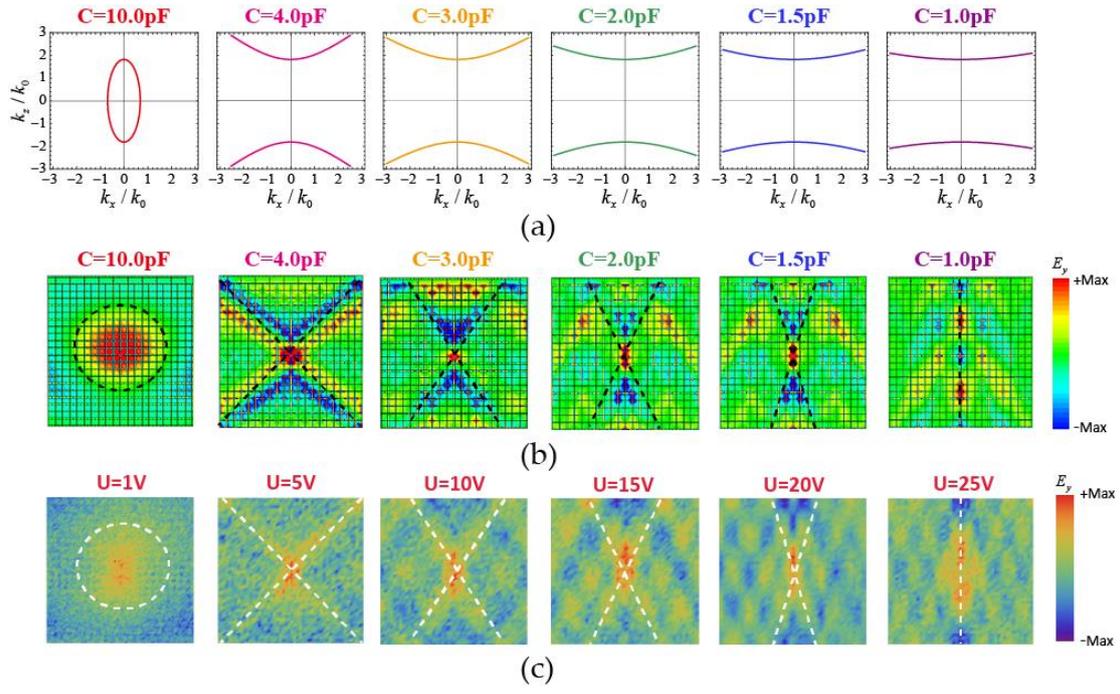